\documentclass[aps,preprint]{revtex4}
\usepackage{amssymb,epsf}
\usepackage{amsmath}
\usepackage{graphicx}
\begin{document}
\title{Dark Energy and Matter in 4 Dimensions From an Empty Kaluza-Klein Spacetime}
\author{M. H. Dehghani$^{1,2}$\footnote{email address:
mhd@shirazu.ac.ir} and Sh. Assyyaee$^{1}$}

\affiliation{$^1$Physics Department and Biruni Observatory,
College of Sciences, Shiraz
University, Shiraz 71454, Iran\\
$^2$Research Institute for Astrophysics and Astronomy of Maragha
(RIAAM), Maragha, Iran}
\begin{abstract}
We consider the third order Lovelock equations without the
cosmological constant term in an empty $n(\geq 8)$-dimensional
Kaluza-Klein spacetime $\mathcal{M}^{4}\times \mathcal{K}^{n-4}$, where $\mathcal{K}^{n-4}$
is a constant curvature space. We show that the emptiness of the higher-dimensional spacetime imposes
a constraint on the metric function(s) of 4-dimensional spacetime $\mathcal{M}^{4}$.
We consider the effects of this
constraint equation in the context of black hole physics, and find a
black hole solution in 4 dimensions in the absence of matter field and the
cosmological constant (dark energy).
This solution has the same form as the 4-dimensional solution
introduced in \cite{Dad1} for Gauss-Bonnet gravity in the presence of cosmological constant,
and therefore the metric of $\mathcal{M}^{4}$ which satisfies the vacuum Lovelock equations
in higher-dimensional Kaluza-Klein spacetime is unique.
This black hole solution shows that
the curvature of an empty higher-dimensional Kaluza-Klein spacetime
creates dark energy and matter with non-traceless energy-momentum tensor in 4 dimensions.
\end{abstract}

\maketitle

\section{Introduction}

High-precision observational data have confirmed with startling evidence
that the universe is undergoing a phase of accelerated expansion \cite{Acc-Exp}.
The cause of this acceleration still remains an open and
tantalizing question. In the standard cosmological model, where the
acceleration of the universe is taken into account by a positive
cosmological constant term, dark energy is responsible for the
acceleration of the universe (see Ref. \cite{Sami} for a review and
references therein). The value of the energy density stored in the
cosmological constant today, has to be of order of the critical density, namely
$10^{-3}$ $eV^{4}$. This value seems arbitrarily small and the known
mechanisms, such as the popular $TeV$-scale supersymmetry breaking scenario
or any top-down high-scale particle physics mechanisms, fail to produce it.

Rather than dealing directly with the cosmological constant, one may also
explore the alternative view point through a modified gravity approach. A
very promising way to explain these major problems is to assume that at
large scales, or higher dimensional spacetime, Einstein theory of General
Relativity breaks down and a more general action should describe the
gravitational field. In this context, an interesting possibility is the
existence of extra dimensions. It is widely believed that string theory is
moving towards a viable quantum gravity theory, and one of the key
predictions of string theory is precisely the existence of extra spatial
dimensions. Extra dimensional models have proven to be very fruitful in
providing new ways of attacking the old problems. In this context, in recent
years, theories with large extra dimensions have received an explosion of
interests. New models such as brane scenarios
\cite{Brane}, large extra dimensions, and
warped extra dimensions have not only revolutionized the Kaluza-Klein
theory \cite{Wes}, but also shed new light on some long-standing problems in particle
physics and cosmology. Interestingly, theories with large extra dimensions
can be even tested by future collider experiments \cite{LHC}.
A natural modification of gravity in higher dimensions is Lovelock gravity, whose Lagrangian
consist of the dimensionally extended Euler densities. This Lagrangian is
obtained by Lovelock as he tried to calculate the most general tensor that
satisfies properties of Einstein's tensor in higher dimensions \cite{Lovelock}.
Since the Lovelock tensor contains derivatives of metrics of
order not higher than second, and the second order
derivative is linear in the field equations, the initial value
problem is well formulated and the evolution of the gravitating system is uniquely defined.

One of the early works of physicists who consider gravity in
higher-dimensional spacetime was the attempt of Kaluza and Klein, who split
general relativity and electrodynamics from an empty 5-dimensional spacetime
\cite{Kaluza}. Kaluza unified not only gravity and electromagnetism, but
also matter and geometry, for the photon appeared in four dimensions as a
manifestation of empty five-dimensional spacetime. Since the theory
of gravity with the assumptions of Einstein in higher dimensions is Lovelock gravity,
it is natural to use the Lovelock field equations of gravity
instead of the Einstein equation in the context of Kaluza-Klein theory.
The action of Lovelock gravity may also be viewed as the
low energy effective action of string theory \cite{string}.
The generation of dark energy from a (super) string effective action
with higher order curvature corrections and a dynamical dilaton has been investigated in \cite{Od}.
Here, we want to consider third order Lovelock gravity without a cosmological
constant term in an $n$-dimensional empty spacetime which is the product of a $4$%
-dimensional Lorentzian manifold $\mathcal{M}^{4}$ with an ($n-4$%
)-dimensional constant curvature manifold $\mathcal{K}^{n-4}$. In the context
of black hole physics, we investigate the problem of having a $4$-dimensional asymptotically anti-de
Sitter (AdS) or de Sitter (dS) charged black hole in the absence of electromagnetic
field and the cosmological constant term in the field equations of gravity.
This idea has been used for the Gauss-Bonnet gravity, but the authors
were forced to keep the cosmological constant in higher dimensions which weakens
the idea of empty higher-dimensional spacetime \cite{Dad1,Dad2}.

The out line of this work is as follows. In Sec. \ref{Kal}, we decompose the
field equations of third order Lovelock gravity in an $n$-dimensional
spacetime which is homomorphic to $\mathcal{M}^{4}\times \mathcal{K}^{n-4}$,
into two equations and introduce a constraint equation for the vacuum solutions of the field
equation. In Sec. \ref{Sol}, we  present a new
4-dimensional asymptotically (A)dS charged black hole solution in the absence of
cosmological constant and electromagnetic field. Finally, we give some concluding remarks.

\section{KALUZA-KLEIN DECOMPOSITION OF BASIC EQUATIONS \label{Kal}}

The vacuum gravitational
field equations of third order Lovelock gravity may be written as:
\begin{equation}
\mathcal{G}_{\mu \nu }=G_{\mu \nu }^{(1)}+\alpha _{2}G_{\mu \nu
}^{(2)}+\alpha _{3}G_{\mu \nu }^{(3)}=0,  \label{Geq}
\end{equation}
where $\alpha _{i}$'s are Lovelock coefficients, $G_{\mu \nu }^{(1)}$ is
just the Einstein tensor, and $G_{\mu \nu }^{(2)}$ and $G_{\mu \nu }^{(3)}$
are the second and third order Lovelock tensors given as
\begin{equation}
G_{\mu \nu }^{(2)}=2(R_{\mu \sigma \kappa \tau }R_{\nu }^{\phantom{\nu}%
\sigma \kappa \tau }-2R_{\mu \rho \nu \sigma }R^{\rho \sigma }-2R_{\mu
\sigma }R_{\phantom{\sigma}\nu }^{\sigma }+RR_{\mu \nu })-\frac{1}{2}%
\mathcal{L}_{2}g_{\mu \nu },  \label{Love2}
\end{equation}
\begin{eqnarray}
G_{\mu \nu }^{(3)} &=&-3(4R^{\tau \rho \sigma \kappa }R_{\sigma \kappa
\lambda \rho }R_{\phantom{\lambda }{\nu \tau \mu}}^{\lambda }-8R_{%
\phantom{\tau \rho}{\lambda \sigma}}^{\tau \rho }R_{\phantom{\sigma
\kappa}{\tau \mu}}^{\sigma \kappa }R_{\phantom{\lambda }{\nu \rho \kappa}%
}^{\lambda }+2R_{\nu }^{\phantom{\nu}{\tau \sigma \kappa}}R_{\sigma \kappa
\lambda \rho }R_{\phantom{\lambda \rho}{\tau \mu}}^{\lambda \rho }  \nonumber \\
&&-R^{\tau \rho \sigma \kappa }R_{\sigma \kappa \tau \rho }R_{\nu \mu }+8R_{%
\phantom{\tau}{\nu \sigma \rho}}^{\tau }R_{\phantom{\sigma \kappa}{\tau \mu}%
}^{\sigma \kappa }R_{\phantom{\rho}\kappa }^{\rho }+8R_{\phantom
{\sigma}{\nu \tau \kappa}}^{\sigma }R_{\phantom {\tau \rho}{\sigma \mu}%
}^{\tau \rho }R_{\phantom{\kappa}{\rho}}^{\kappa }  \nonumber \\
&&+4R_{\nu }^{\phantom{\nu}{\tau \sigma \kappa}}R_{\sigma \kappa \mu \rho
}R_{\phantom{\rho}{\tau}}^{\rho }-4R_{\nu }^{\phantom{\nu}{\tau \sigma
\kappa }}R_{\sigma \kappa \tau \rho }R_{\phantom{\rho}{\mu}}^{\rho
}+4R^{\tau \rho \sigma \kappa }R_{\sigma \kappa \tau \mu }R_{\nu \rho
}+2RR_{\nu }^{\phantom{\nu}{\kappa \tau \rho}}R_{\tau \rho \kappa \mu }
\nonumber \\
&&+8R_{\phantom{\tau}{\nu \mu \rho }}^{\tau }R_{\phantom{\rho}{\sigma}%
}^{\rho }R_{\phantom{\sigma}{\tau}}^{\sigma }-8R_{\phantom{\sigma}{\nu \tau
\rho }}^{\sigma }R_{\phantom{\tau}{\sigma}}^{\tau }R_{\mu }^{\rho }-8R_{%
\phantom{\tau }{\sigma \mu}}^{\tau \rho }R_{\phantom{\sigma}{\tau }}^{\sigma
}R_{\nu \rho }-4RR_{\phantom{\tau}{\nu \mu \rho }}^{\tau }R_{\phantom{\rho}%
\tau }^{\rho }  \nonumber \\
&&+4R^{\tau \rho }R_{\rho \tau }R_{\nu \mu }-8R_{\phantom{\tau}{\nu}}^{\tau
}R_{\tau \rho }R_{\phantom{\rho}{\mu}}^{\rho }+4RR_{\nu \rho }R_{%
\phantom{\rho}{\mu }}^{\rho }-R^{2}R_{\nu \mu })-\frac{1}{2}\mathcal{L}%
_{3}g_{\mu \nu }.  \label{Love3}
\end{eqnarray}
In Eqs. (\ref{Love2}) and (\ref{Love3}) $\mathcal{L}_{2}=R_{\mu \nu \gamma
\delta }R^{\mu \nu \gamma \delta }-4R_{\mu \nu }R^{\mu \nu }+R^{2}$ is the
Gauss-Bonnet Lagrangian and
\begin{eqnarray}
\mathcal{L}_{3} &=&2R^{\mu \nu \sigma \kappa }R_{\sigma \kappa \rho \tau }R_{%
\phantom{\rho \tau }{\mu \nu }}^{\rho \tau }+8R_{\phantom{\mu \nu}{\sigma
\rho}}^{\mu \nu }R_{\phantom {\sigma \kappa} {\nu \tau}}^{\sigma \kappa }R_{%
\phantom{\rho \tau}{ \mu \kappa}}^{\rho \tau }+24R^{\mu \nu \sigma \kappa
}R_{\sigma \kappa \nu \rho }R_{\phantom{\rho}{\mu}}^{\rho }  \nonumber \\
&&+3RR^{\mu \nu \sigma \kappa }R_{\sigma \kappa \mu \nu }+24R^{\mu \nu
\sigma \kappa }R_{\sigma \mu }R_{\kappa \nu }+16R^{\mu \nu }R_{\nu \sigma
}R_{\phantom{\sigma}{\mu}}^{\sigma }-12RR^{\mu \nu }R_{\mu \nu }+R^{3}
\label{L3}
\end{eqnarray}
is the third order Lovelock Lagrangian. Equation (\ref{Geq}) does not
contain the derivative of the curvatures, and therefore the derivatives of
the metric higher than two do not appear. In order to have the contribution
of all the above terms in the field equation, the dimension of the spacetime
should be equal or larger than seven. Solutions
of third order Lovelock gravity have been introduced in \cite{Deh}. Here, we
investigate the Kaluza-Klein solutions of third order Lovelock gravity.

Consider an $n$-dimensional spacetime which is homomorphic to $\mathcal{M}%
^{4}\times \mathcal{K}^{n-4}$ for $n\geq 8$, with the metric
\begin{equation}
ds^{2}=g_{ab}dx^{a}dx^{b}+r_{0}^{2}\gamma _{ij}d\theta ^{i}d\theta ^{j},
\label{met}
\end{equation}
where $a,b=0,1,2,3$ and $i,j=4...(n-1)$. In Eq. (\ref{met}) $g_{ab}$ is a
Lorentzian metric on $\mathcal{M}^{4}$, $r_{0}$ is a constant and $\gamma_{ij}$
is the metric on the $(n-4)$-dimensional constant curvature space $\mathcal{K}^{n-4}$ with
curvature $(n-4)(n-5)k$, where $k=0,\pm 1$. It is a matter
of calculation to show that the tensor $\mathcal{G}_{\,\,\,\,\nu }^{\mu }$
gets the below decomposition
\begin{eqnarray}
\mathcal{G}_{\ b}^{a} &=&-\frac{(n-4)(n-5)k}{2r_{0}^{2}}\left\{ 1+\frac{k\alpha _{2}(n-6)(n-7)}{%
r_{0}^{2}}+\frac{\alpha _{3}(n-6)(n-7)(n-8)(n-9)}{r_{0}^{4}}\right\} \delta
_{\ b}^{a}  \nonumber \\
&& +\left\{ 1+\frac{2k\alpha _{2}(n-4)(n-5)}{r_{0}^{2}}+%
\frac{3\alpha _{3}k^{2}(n-4)(n-5)(n-6)(n-7)}{r_{0}^{4}}\right\} \hat{G}_{\
b}^{a},  \label{Gab}
\end{eqnarray}
\begin{eqnarray}
\mathcal{G}_{\ j}^{i} &=&-\frac{1}{2} \left\{\frac{(n-5)(n-6)k}{r_{0}^{2}}\left[ 1+\frac{k\alpha _{2}(n-7)(n-8)}{%
r_{0}^{2}}+\frac{k^{2}\alpha _{3}(n-7)(n-8)(n-9)(n-10)}{r_{0}^{4}}\right] \right. \nonumber \\
&&\hspace{1cm} \left.+\left[ 1+\frac{2k\alpha _{2}(n-5)(n-6)}{r_{0}^{2}}+\frac{3k^{2}\alpha
_{3}(n-5)(n-6)(n-7)(n-8)}{r_{0}^{4}}\right]\right. \hat{R}  \nonumber \\
&& \hspace{1cm} \left.+\left[ \alpha _{2}+%
\frac{3k\alpha _{3}(n-5)(n-6)}{r_{0}^{2}}\right] \hat{L}_{2} \right\}\delta _{\ j}^{i}
\label{Gij}
\end{eqnarray}
where the superscripts ``hat'' represent quantities on $\mathcal{M}^{4}$.

Now as in the case of Kaluza-Klein theory, we want to obtain the vacuum
solutions of Eqs. (\ref{Gab}) and (\ref{Gij}). In general, $\hat{G}_{\ b}^{a}\neq 0$
(as we will see in the next section),
and therefore $\mathcal{G}_{\,\,\,\,b}^{a}=0$ if the two brackets in Eq. (\ref{Gab}%
) vanish, which concludes that $k\neq0$. For $k=\pm 1$, one obtains:
\begin{eqnarray}
&& \alpha _{2}=-\frac{2k(n-6)(n+1)}{(n-4)(n-5)(n^{2}-5n-18)}r_{0}^{2},
\label{A2}\\
&& \alpha _{3}=\frac{(n^{2}-5n-2)}{(n-4)(n-5)(n-6)(n-7)(n^{2}-5n-18)}r_{0}^{4}.
\label{A3}
\end{eqnarray}
Note that $n$ should be greater than 7 ($n\geq 8$) in order to have a finite
value for $\alpha _{3}$. It is worthwhile to mention that the Lovelock
coefficients $\alpha _{2}$ and $\alpha _{3}$ are proportional to $r_{0}^{2}$
and $r_{0}^{4}$, respectively, where $r_0$ is the size of the extra dimensions.
Thus, the smallness of $r_{0}$ is consistent with the fact that
the Lovelock coefficients are supposed to be very small. Substituting
the above $\alpha _{2}$ and $\alpha _{3}$ in Eq. (\ref{Gij}), one obtains:
\begin{equation}
r_{0}^{4}\hat{L}_{2}-kA_{n}r_{0}^{2}\hat{R}+B_{n}=0,  \label{Cons}
\end{equation}
where
\begin{eqnarray}
A_{n} &=&\frac{4(n-2)(n-3)(n-5)(n-7)}{n^{3}-6n^{2}+11n-54}, \\
B_{n} &=&\frac{8(n-5)(n-7)(2n^{3}-27n^{2}+97n-54)}{%
n^{3}-6n^{2}+11n-54}.
\end{eqnarray}
Equation (\ref{Cons}) should be used as a constraint on the metric
function(s) of the 4-dimensional spacetime $\mathcal{M}^{4}$.

It is worth noting that one needs two parameters in Eq. (\ref{Gab})
in order to have $G^a_{\ b}=0$. In Ref. \cite{Dad1}, the two parameters
have been chosen to be the cosmological constant
and Gauss-Bonnet coefficient, while in this paper we choose them
to be the second and third order Lovelock coefficients.
The presence of the cosmological constant in Ref. \cite{Dad1}
weakens the idea of emptiness of higher-dimensional Kaluza-Klein spacetime.
Of course, one may use any order of Lovelock gravity with at least two parameters,
but in any order of Lovelock gravity the constraint equation (\ref{Cons})
for $\mathcal{M}^{4}\times \mathcal{K}^{n-4}$ has the same form, and therefore the
4-dimensional solution which satisfies the vacuum Lovelock equations is unique.
\section{Four-Dimensional Black Hole Solutions\label{Sol}}

In this section, we investigate the effects of the constraint equation (\ref{Cons})
in the context of black hole physics. In
order to do this, we use the metric of spherically symmetric spacetime in
the Schwarzschild gauge, $g_{tt}g_{rr}=-1$, for $\mathcal{M}^{4}$ manifold:
\begin{equation}
g_{ab}dx^{a}dx^{b}=-f(r)dt^{2}+\frac{1}{f(r)}dr^{2}+r^{2}d\Sigma _{2(\hat{k}%
)}^{2},  \label{met3}
\end{equation}
where $d\Sigma _{2(\hat{k})}^{2}$ represents the line element of a $2$%
-dimensional hypersurface with constant curvature $2\hat{k}$ and $\hat{k}%
=0,\pm 1$. One can show that Eq. (\ref{Cons}) for the above metric after
integrating two times reduces to
\begin{equation}
2r_{0}^{4}\left( f(r)-\hat{k}\right) ^{2}+kA_{n}r_{0}^{2}r^{2}\left( f(r)-%
\hat{k}\right) +\frac{B_{n}}{12}r^{4}-C_{1}r+C_{2}=0, \label{Eqfr}
\end{equation}
where $C_{1}$ and $C_{2}$ are two arbitrary constants of integration. Thus, the metric function $f(r)$
may be written as:
\begin{equation}
f(r)=\hat{k}-A_{n}\frac{r^{2}}{4r_{0}^{2}}\ \left( k\pm \sqrt{1-\frac{%
2B_{n}}{3A_{n}^{2}}+\frac{8C_{1}}{A_{n}^{2}r^{3}}-\frac{8C_{2}}{%
A_{n}^{2}r^{4}}}\right),  \label{fr1}
\end{equation}
which has the same form as the metric function introduced by Maeda and Dadhich,
and therefore it establishes the uniqueness of Maeda-Dadhich black hole solution \cite{Dad1}.

The function $f(r)$ is
real provided $r\geq r_{\min }$, where $r_{\min }$ is the largest real root
of the square root:
\begin{equation}
(3A_{n}^{2}-2B_{n})r_{\min }^{4}+24C_{1}r_{\min }-24C_{2}=0.
\end{equation}
For negative values of $C_2$,
the above equation has no real root and therefore the metric function is
real for all the spacetime. But for positive values of $C_2$,
for which the solution is asymptotically Reissner-Nordstrom black hole as we will see below,
one may restrict the spacetime to the region $r\geq r_{\min }$
by the transformation $\rho^2=r^2-r_{\min }^2$ (see the Appendix for more details).

In order to study the general structure of this solution, we first look for
the curvature singularities. It is easy to show that the Kretschmann scalar $%
R_{\mu \nu \lambda \kappa }R^{\mu \nu \lambda \kappa }$ diverges at $r=0$
or  $r=r_{\min }$ for negative or positive values of $C_2$, respectively.
Here, we are interested in black hole
solutions, and therefore we only consider the plus branch of Eq. (\ref{fr1}).
The minus branch presents a naked singularity.
Seeking possible black hole solutions, we turn to
looking for the existence of horizons. The event horizon(s), if there exists
any, is (are) located at the root(s) of $f(r)=0$:
\begin{equation}
\frac{B_{n}}{12}r_{+}^{4}-k\hat{k}%
A_{n}r_{0}^{2}r_{+}^{2}-C_{1}r_{+}+C_{2}+2r_{0}^{4}=0.  \label{Hor}
\end{equation}
Equation (\ref{Hor}) may have zero, one or two real positive solutions
depending on the suitable choices of $C_{1}$and $C_{2}$. Thus, the
topological solution given by Eqs. (\ref{met3}) and (\ref{fr1}) may be
interpreted as a naked singularity, an extreme black hole or a black
hole with two inner and outer event horizons.
For extreme black hole, both $f(r)$ and $f^{\prime }(r)$ are zero at the
horizon $r=r_{\mathrm{ext}}$. Differentiating Eq. (\ref{Eqfr}) with respect
to $r$ and using $f(r_{\mathrm{ext}})=f^{\prime }(r_{\mathrm{ext}})=0$, it
is easy to show that:
\begin{equation}
B_{n}r_{\mathrm{ext}}^{3}-6k\hat{k}r_{0}^{2}A_{n}r_{\mathrm{ext}%
}-3C_{1}=0.  \label{rext}
\end{equation}
Using Eq. (\ref{Hor}) and (\ref{rext}) for $r_{\mathrm{ext}}$, one can show
that the solution (\ref{fr1}) present an extreme black hole provided $%
C_{1}=C_{\mathrm{ext}}$, has two inner and outer horizons provided $C_{1}>C_{%
\mathrm{ext}}$, and a naked singularity if $C_{1}<C_{1\mathrm{ext}}$, where $%
C_{1\mathrm{ext}}$ is
\begin{equation}
C_{1\mathrm{ext}}=\left( \sqrt{A_{n}^{2}r_{0}^{4}+B_{n}(C_{2}+2r_{0}^{4})}-2k%
\hat{k}r_{0}^{2}A_{n}\right) \left( \frac{8k\hat{k}r_{0}^{2}A_{n}+8\sqrt{%
A_{n}^{2}r_{0}^{4}+B_{n}(C_{2}+2r_{0}^{4})}}{9B_{n}}\right) ^{1/2}.
\end{equation}
\begin{figure}
\centering {\includegraphics[width=5cm]{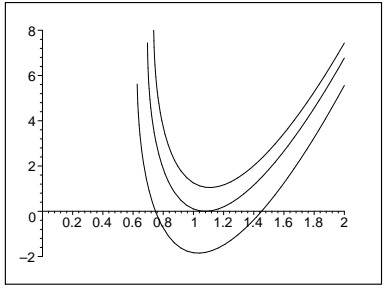} }
\caption{$f(r)$ versus $r$ for $n=8$, $k=-1$, $\hat{k}=1$, $C_2 =1.0$, $r_{0}=.2$,
$C_1<C_{1\mathrm{ext}}$, $C_1=C_{1\mathrm{ext}}$ and $C_1>C_{1\mathrm{ext}}$ from up to down, respectively.} \label{Fr}
\end{figure}
Figure \ref{Fr} shows the function $f(r)$ versus $r$ for different
values of $C_1$.

It is a matter of calculation to show that the asymptotic behavior of the
solution is (A)dS with the effective cosmological constant
\begin{equation}
\Lambda _{\mathrm{eff}}=\frac{3A_{n}}{4r_{0}^{2}}\left( k+\sqrt{1-\frac{%
2B_{n}}{3A_{n}^{2}}}\right) .  \label{Leff}
\end{equation}
Thus, the asymptotic behavior of the solution is dS for $k=1$, and is AdS for $k=-1$.
The metric function $f(r)$ at large value of $r$ may be written
as
\begin{equation}
f(r)=\hat{k}-\frac{\Lambda _{\mathrm{eff}}}{3}r^{2}-\frac{2m}{r}+\frac{q^{2}}{r^{2}}%
,
\end{equation}
where
\begin{eqnarray}
m &=&\frac{C_{1}}{2r_{0}^{2}}\sqrt{1-\frac{2B_{n}}{3A_{n}^{2}}}, \label{mass}\\
q^{2} &=&\frac{C_{2}}{r_{0}^{2}}\sqrt{1-\frac{2B_{n}}{3A_{n}^{2}}}\label{charge}.
\end{eqnarray}
Thus, the above solution at large values of $r$ behaves as the
Reissner-Nordstorm-(A)dS black hole in the absence of cosmological constant
and any kind of electromagnetic field provided $C_{2}>0$. The parameters $C_{1}$ and $C_{2}$
may be related to the mass and charge parameters of the spacetime according
to Eqs. (\ref{mass}) and (\ref{charge}).
\begin{figure}
\centering {\includegraphics[width=7cm]{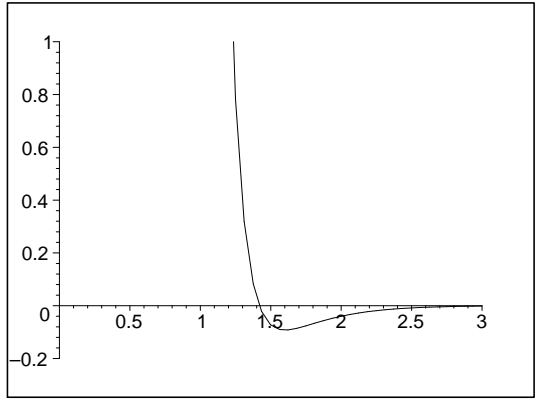} }
\caption{Trace of energy-momentum of matter versus $r$ for $n=8$, $k=-1$, $\hat{k}=1$, $C_1 =0.5$, $C_2=.6$, $r_{0}=.2$.} \label{Trace}
\end{figure}

In order to investigate the nature of the matter field created by the curvature of
the higher-dimensional Kaluza-Klein spacetime in 4 dimensions, we first set the arbitrary
integration constants $C_1$ and $C_2$ equal to zero. In the absence of matter ($C_1=C_2=0$), the 4-dimensional
Einstein tensor $\hat{G}_{ab}$ will be proportional
to $g_{ab}$, where the constant of proportionality is the effective cosmological constant.
This fact persuades us to split the effective
4-dimensional energy-momentum tensor created by the curvature of
the higher-dimensional Kaluza-Klein spacetime into
a part which is due to the effective cosmological constant and the reminder of it as follows:
\begin{eqnarray}
T_{ab}^{\mathrm{DE}} &=&-\frac{\Lambda _{\mathrm{eff}}}{8\pi}g_{ab}, \\
T_{ab}^{\mathrm{M}} &=&\frac{1}{8 \pi} \left( \hat{G}_{ab}+\Lambda _{\mathrm{eff}}g_{ab}\right),
\end{eqnarray}
where the superscripts ``DE'' and ``M'' used for the abbreviation of dark
energy and matter, respectively. $T_{ab}^{\mathrm{DE}}$ is the energy-momentum
tensor of the effective cosmological constant, and $T_{ab}^{\mathrm{M}}$ is the
energy-momentum tensor of the matter created by the curvature of the
higher-dimensional Kaluza-Klein spacetime. The word ``dark" is used
to emphasize that the origin of dark energy is the
curvature of higher dimensional Kaluza-Klein spacetime in Lovelock gravity.
Note that at infinity, the effect of matter is vanished, and we leave with the
energy-momentum of the dark energy. It is a matter of calculations to
show that the radial and tangential pressures of
the matter part at large $r$ reduce to
\begin{eqnarray}
P^M_r&=&-\frac{C_2}{8\pi A_n r_0^2} \left({1-\frac{2B_n}{3A_n}}\right)^{-1/2}\frac{1}{r^4}-\frac{6C_1^2}{8\pi A_n^3 r_0^2} \left({1-\frac{2B_n}{3A_n}}\right)^{-3/2}\frac{1}{r^6}+\mathcal{O}\left(\frac{1}{r^8}\right), \\
P^M_t&=&\frac{C_2}{8\pi A_n r_0^2} \left({1-\frac{2B_n}{3A_n}}\right)^{-1/2}\frac{1}{r^4}
+\frac{12C_1^2}{8\pi A_n^3 r_0^2} \left({1-\frac{2B_n}{3A_n}}\right)^{-3/2}\frac{1}{r^6}
+\mathcal{O}\left(\frac{1}{r^8}\right),
\end{eqnarray}
respectively, which are sufficiently small at large $r$.
Also, it is worth to note that $T_{t}^{t\mathrm{(M)}}=T_{r}^{r\mathrm{(M)}}$and
$T_{\theta}^{\theta \mathrm{(M)}}=T_{\varphi }^{\varphi \mathrm{(M)}}$, but the energy
momentum tensor $T_{ab}^{\mathrm{M}}$ is not traceless at finite $r$.
This can be seen from Eq. (\ref{Cons}) which shows that $\hat{R}$ is
a linear combination of $\hat{L}_2$ and $B_n$ and therefore
is a function of $r$, that reduces to a constant as $r$ goes to infinity.
Figure \ref{Trace} shows the trace of energy-momentum of matter versus
$r$, which shows that it becomes traceless at infinity. Thus,
the electromagnetic field which can be related to this solution is not
Maxwellian.

The solution presented in this subsection shows the creation of dark energy
and matter out of pure curvature of a higher-dimensional Kaluza-Klein
spacetime.

\section{Closing Remarks}
We decomposed the field equations of third order Lovelock
gravity in the absence of cosmological constant and matter field in an $n$%
-dimensional Kaluza-Klein spacetime, $\mathcal{M}^{4}\times \mathcal{K}%
^{n-4} $, into two equations. One of these equations fixes the Lovelock
coefficients in terms of the size of the extra dimensions $r_0$ and the dimension of the
Kaluza-Klein spacetime, and the second one should be used as a constraint on
the four dimensional metric of $\mathcal{M}^{4}$. The proportionality of the
Lovelock coefficients and the powers of $r_0$ [relations (\ref{A2}) and (\ref
{A3})] shows that the smallness of $r_{0}$, which is needed for completing
the scenario of Kaluza-Klein, is consistent with the smallness of Lovelock
coupling constant.

We performed our formalism in the context of black hole
physics, and found an asymptotically (A)dS charged black
hole solution in four dimensions which behaves like Reissner-Nordtrom-(A)dS
black hole at large distance from the singularity of the spacetime. In this
scenario, we found that the curvature of higher-dimensional Kaluza-Klein
spacetime creates the effects of matter and dark energy in four dimensions.
That is, one may have asymptotically (A)dS charged black hole solutions in
the absence of the cosmological constant and electromagnetic field. We also
found that the electromagnetic field is not Maxwellian. This solution
is the same as the solution which is introduced by Maeda and Dadhich, and
therefore it establishes the uniqueness of Maeda-Dadhich black hole solution \cite{Dad1}.

Now, one may ask about the sources of $q$ and $\Lambda_{\mathrm{eff}}$.
Indeed, the Ricci scalar in an empty 4-dimensional spacetime should be zero,
while we found that $\hat{R}$ does not vanish, and depends on the coordinate $r$.
At large $r$, both $\hat{R}$ and $\hat{L}_2$ become constant and therefore the source of $q$ is
asymptotically Maxwellian, while at finite $r$, the matter created by the
curvature of higher-dimensional Kaluza-Klein spacetime is not traceless. It is worthwhile to mention that
this creation of dark energy and matter out of pure curvature of a
higher-dimensional Kaluza-Klein spacetime is the generalization of the
creation of matter out of pure curvature discussed in Ref. \cite{Dad1}.

\textbf{Acknowledgements} This work has been supported by Research Institute
for Astrophysics and Astronomy of Maragha.

\begin{center}
\large{\textbf{APPENDIX}}
\end{center}

In order to restrict the spacetime to the
physical region $r \geq r_{\mathrm{min}}$, we introduce a new radial coordinate $\rho$ as:
\begin{equation*}
\rho^{2}=r^{2}-r_{\mathrm{min}}^{2}\Rightarrow dr ^{2}=\frac{\rho^{2}}{\rho^{2}+r_{\mathrm{min}}^{2}}%
d\rho^{2},
\end{equation*}
where now $\rho$ is in the range $0\leq \rho <\infty$.
With this new coordinate, the metric (\ref{met3}) becomes:
\begin{equation*}
g_{ab}dx^{a}dx^{b}=-f(\rho)dt^{2}+\frac{\rho^2d\rho^{2}}{(\rho^2+r_{\mathrm{min}}^{2})f(\rho)}+(\rho^2+r_{\mathrm{min}}^{2})d\Sigma _{2(\hat{k}%
)}^{2},
\end{equation*}
where now $f(\rho)$ is
\begin{equation*}
f(\rho)=\hat{k}-A_{n}\frac{\rho^2+r_{\mathrm{min}}^{2}}{4r_{0}^{2}} \left( k\pm \sqrt{1-\frac{%
2B_{n}}{3A_{n}^{2}}+\frac{8C_{1}}{A_{n}^{2}(\rho^2+r_{\mathrm{min}}^{2})^{3/2}}-\frac{8C_{2}}{%
A_{n}^{2}(\rho^2+r_{\mathrm{min}}^{2})^{2}}}\right).  \label{fr}
\end{equation*}


\begin{thebibliography}{9}
\bibitem{Acc-Exp}  A.G. Riess et al., Astron. J. 516 (1998) 1009;\\
S. Perlmutter et al., Astron. J. 517 (1999) 565;\\
S. Perlmutter, M.S. Turner and M. White, Phys. Rev. Lett. 83 (1999) 670;\\
A. Grant et al., Astron. J. 560 (2001) 49;\\
A. G. Riess et al., Astron. J. 607 (2004) 665;\\

\bibitem{Sami}  E. J. Copeland, M. Sami, and S. Tsujikawa, Int. J. Mod.
Phys. D 15 (2006) 1753.

\bibitem{Brane} L. Randall and R. Sundrum, Phys. Rev. Lett. 83 (1999) 3370; 4690;\\
N. Arkani-Hamed, S. Dimopoulos, and G. Dvali, Phys. Lett. B 429 (1998) 263;\\
I. Antoniadis, N. Arkani-Hamed, S. Dimopoulos, and G. Dvali, Phys. Lett. B 436 (1998) 257;\\
G. Dvali, G. Gabadadze, and M. Porrati, Phys. Lett. B 485 (2000) 208;\\
G. Dvali and G. Gabadadze, Phys. Rev. D 63 (2001) 065007;\\
G. Dvali, G. Gabadadze, and M. Shifman, Phys. Rev. D 67 (2003) 044020.

\bibitem{Wes} J. M. Overduin and P. S. Wesson, Phys. Rept. 283 (1997) 303.

\bibitem{LHC}  S. Dimopoulos, G. Landsberg, Phys. Rev. Lett. 87 (2001) 161602;\\
A. Chamblin and G.C. Nayak, Phys. Rev. D 66 (2002) 091901;\\
S.B. Giddings and S. Thomas, Phys. Rev. D 65 (2002) 056010;\\
P. Kanti, Int. J. Mod. Phys. A 19 (2004) 4899;\\
D.M. Gingrich, Int. J. Mod. Phys. A 21 (2006) 6653 (2006);\\
G. Landsberg, J. Phys. G 32 (2006) R337.

\bibitem{Lovelock}  D. Lovelock, J. Math. Phys. 12 (1971) 498.

\bibitem{Kaluza}  Th. Kaluza, Sitz. Preuss. Akad. Wiss. Leipzig (1921) 966;\\
O. Klein, Zeits. Phys. 37 (1926) 895.

\bibitem{string} M. B. Greens, J. H. Schwarz, and E. Witten, Superstring
Theory (Cambridge University Press, Cambridge,
England, 1987).

\bibitem{Od} S. Nojiri, S. D. Odintsov and M. Sami,  Phys. Rev. D 74 (2006) 046004;\\
E. Elizaldel et. al., Eur. Phys. J. C 53 (2008) 447.

\bibitem{Dad1}  H. Maeda and N. Dadhich, Phys. Rev. D 74 (2006) 021501(R).
\bibitem{Dad2} H. Maeda and N. Dadhich, Phys. Rev. D 75 (2007) 044007;\\
N. Dadhich and H. Maeda, Int. J. Mod. Phys. D 70 (2008) 513;\\
A. Molina and N. Dadhich, arXiv:08041194, Int. J. Mod. Phys. D, in press.

\bibitem{Deh} M.H. Dehghani and M. Shamirzaie, Phys. Rev. D 72 (2005) 124015;\\
M.H. Dehghani and R.B. Mann, Rev. D 73 (2006) 104003;\\
M.H. Dehghani and N. Farhangkhah, Phys. Lett. B 674 (2009) 243.

\end{thebibliography}
\end{document}